\documentclass[12pt]{article}
\usepackage{amssymb}
\usepackage{graphicx}
\usepackage{amsmath}
\usepackage{float}
\usepackage{ragged2e}
\usepackage{multirow}
\usepackage{titlesec}

\titleformat{\paragraph}
{\normalfont\normalsize\bfseries}{\theparagraph}{1em}{}
\titlespacing*{\paragraph}
{0pt}{3.25ex plus 1ex minus .2ex}{1.5ex plus .2ex}

\begin{document}

\newpage
\pagestyle{plain}
\setcounter{page}{1}

\centering{\fontsize{18}{9}\bf{Constructing Social Networks From Binary Data}}\\ \ \\
\centering{\fontsize{15}{9}{Sirui Wang and Mei Wang  \\
The University of Chicago}}\\
\centering{\fontsize{12}{9}{June 8, 2018}}

\justify

\begin{abstract}
\vspace{4mm}\noindent Much of applied network analysis concerns with studying the existing relationships between a set of agents; however, little focus has been given to the considerations of how to represent observed phenomena as a network object. In the case of physical structures such as electric grids or transportation flows, the construction of a network model is fairly straightforward as the nodes and edges usually correspond to some physical structure themselves. On the other hand, construction of a social network is much less defined; while nodes may correspond to well defined social agents, such as people or groups of people, there much more liberty in defining the relationship that edges should represent. This paper studies the intricacies of constructing a social network from data, in particular, binary data, applicable to a wide range of social science contexts. We examine several methods of constructing social networks in prior literature and discuss the methods under a common framework. Finally, using a data set of meetings among technologists in New York, we show that the different constructions of social networks arising from various interpretations of the underlying social relationships can result in vastly different network structures. These findings highlight the significance of understanding the precise relationships of interest when building a network model.
\end{abstract}

\tableofcontents
\newpage

\section{Introduction}

\vspace{4mm}\noindent Network analysis has become a popular way to study physical structures (e.g. the Internet, power grids and transportation systems), information (e.g. the World Wide Web), patents and citations, and social interactions. While physical and information networks have fairly intuitive links between their agents to form edges of the network, social networks ties are usually defined by some form of social relationship, and ``the particular definition one uses will depend on what questions one is interested in answering" \cite{newman}. Social ties often focus on agents' similarities, relations, interactions or flows \cite{borgatti}. Unlike physical networks, social networks are often based on open systems where the set of nodes and edges are not fixed, and inference from the network are generally meant to generalize to a broader population, or the same population repeated interacting over time. Kadushin (2012) suggests that social concepts such as propinquity and homophily affect how \textit{likely} nodes with particular attributes would form ties in a social network, accentuating the idea that social ties are more appropriately described as \text{probabilistic} and \text{dynamic}, rather than \text{deterministic} and \text{static} as one would imagine in physical networks \cite{kadushin}. Accordingly, social networks also easily lend themselves to a modeling approach of network analysis.

\vspace{4mm}\noindent Often times in empirical settings, researchers collect data on a binary outcome from a collection of agents and wish to study the underlying relationships between the agents through the data. This is especially common in social science studies interested in similarity or social relations ties, such as determining voting or purchasing patterns among individuals or determining links between subgroups of a community where individuals maintain their own set of group memberships (e.g. \cite{porter},\cite{kartik},\cite{dellaposta}). A network, therefore, is an attractive way to visualize and analyze the relationships between the agents in a study; this representation of the underlying relationships between agents can also lend itself to a multitude of network-specific methods for inference, prediction or community detection. In these contexts, however, similarities or relations can take on a more ambiguous interpretation than traditional physical or flow networks.

\vspace{4mm}\noindent In a study of the relationships between agents through a series of their actions, one way to define a social network is to have an edge between two nodes (e.g. politicians) represent an association with respect to the binary outcome (e.g. similar voting patterns). While the interpretation of the network nodes are fairly straight-forward, there is liberty in how the edges can be defined, which can affect interpretation and further inference. For example, a study can be designed to examine the voting behavior of a group of politicians based on their voting records on past bills to which each politician voted either ``yes" or ``no". Define each node in the network to represent a politician and draw an edge between two politicians, A and B, if they both ``tended to agree" in their voting decisions. This gives an intuitive and valid network representation of the data and allows for various network-based learning methods. But how can ``tending to agree" be formalized in technical sense? We may want to define this as having the proportion of the two politicians agreeing on the bill to be above a certain threshold. We many alternatively want to define this as if knowing how A voted allows us to guess more accurately on how B voted. These formalizations are both valid, yet conceptually different, ways of constructing a network representation from the data. This simple example illustrates that subtleties in the construction of the network can lead to differing aspects of inference and suggests that scientists who wish to use network-based methods on their data should give some consideration on which construction of the network best answers the question of interest. This paper surveys some common ways of define and construct the network using binary data and discusses the differences in each method.

\vspace{4mm}\noindent Some empirical studies treat a social network as one observation of some underlying phenomenon; these studies construct a network from a data set and calculate various network statistics of interest from the constructed network (e.g. \cite{aral}, \cite{susarla}). When constructing a network from data, edges would often be drawn using some criteria based on co-ocurrence or correlation. Inference from these studies are generally not done on the network statistics themselves, but on secondary parameters that the network characteristics inform (such as used in a regression). In other studies, networks are constructed from data to make inference on underlying phenomenon that can be responsible for future observations (e.g. \cite{porter}). These studies treat the network as a generative model. In this framework, the data collected can be viewed as realizations of the underlying phenomenon infused with some noise; the problem is then to estimate a model that is believed to have generated the data. Since the network edges intuitively signal ``relatedness" or ``association", the network model should take into account underlying \textit{probabilities} of co-occurence in its edges. In the voting example, we may hope to estimate a generative model where every node is a politician and an edge represents a ``high" probability that the two politicians would vote the same way. This is analogous to the case where we want an edge to reflect how often the politicians agree with each other. The alternative case where knowing how one politician votes on the bill reveals some information on another would vote relates to the probabilisitic concept of \textit{independence}. A generative model would allow the observed data phenomenon to be predictive of non-observed events.

\vspace{4mm}\noindent The following sections of this paper considers both non-model based and model based approaches for estimating social ties, an empirical example and discussion. Section 2 addresses using co-occurence probability (proportions for non-model based) and marginal independence (correlation) as ways to define network edges and how each method leads to subtle differences in interpretation of the estimated network. Section 3 discusses how the non-model based approaches are a special case of dyadic independence model for network ties and gives a brief discussion on a more sophisticated exponential random graph model (ERGM) based on partial conditional independence between ties. Section 4 introduces a data set of memberships in technology-based Meetup groups in New York hosted on Meetup.com. The structure of the data set lends itself to a network model, and we apply the different methods of network construction and demonstrate the potential insights each method can extract from this type of data. Section 5 presents a discussion on the importance of the differences in interpretations provided by each network estimation method and the strengths and limitations in inference that each can provide.

\section{Empirical Network Construction}

\vspace{4mm}\noindent Edges of a network represent an association between two nodes and can be determined by a quantification of association between two nodes. Two nodes that are not connected with an edge can, therefore, be interpreted as having negligible association with respect to the quantification. In a non-model based approach to network analysis, the objective is to construct a single network from observed data. If $X$ and $Y$ are nodes in the network whose binary outcomes are observed, the relationship between the two agents can be measured through co-occurrence counts and correlation measures of association, both of which can be characterized by the proportions presented in Table \ref{probs}.  
\begin{table}[h!]
\caption{Observed proportions of $X$ and $Y$; $0\leq p_{ij} \leq 1,\sum_{i,j}p_{ij} = 1$.}

\centering
\begin{tabular}{*4c} 
 & &\multicolumn{2}{c}{$Y$}\\
 & & $0$ &$1$ \\\cline{3-4}
\multirow{2}{*}{$X$}& $0$ & \multicolumn{1}{|c|}{$p_{00}$} & \multicolumn{1}{c|}{$p_{01}$} \\\cline{3-4}
& $1$ & \multicolumn{1}{|c|}{$p_{10}$} &\multicolumn{1}{c|}{$p_{11}$}\\\cline{3-4}
\end{tabular}
\label{probs}
\end{table}

\vspace{4mm}\noindent $p_{00}$ represents the proportion when the outcomes $X$ and $Y$ are both 0; $p_{01}$ represents the proportion when $X$ is 1 and $Y$ is 0, and so on. The raw counts for each cell can be retrieved by simply multiplying the respective proportion by $N$, the sample size of the observed data. This section explores a number of potential quantities that measure different aspects of association; each measure can lead to a reasonable network representation of binary data but have subtle differences in interpretation. Outcomes of the agents $X$ and $Y$ may be referred to variables in following subsections.

\subsection{Co-occurrence}

\vspace{4mm}\noindent Co-occurrence is perhaps the most intuitive way to capture association. In a symmetric setting, a measure of co-occurrence may be $p_{00}+p_{11}$, which is the proportion that the outcomes of the two agents are the same. 

\vspace{4mm}\noindent There may be circumstances, however, when either $0$ or $1$ is the norm, and agents that agree on one value is not as contextually significant as if they agreed on the other. In that case, simply $p_{00}$ or $p_{11}$ instead of their sum may be an adequate measure of co-occurrence. For instance, if the network of interest is a social network where the nodes are students and edges are weighted by how similar a pair of students' interests are measured based on the clubs they join, a club in which 2 students are both students take part in reveals more about their interests than do a club in which neither student take part reveal about their disinterests, especially if the number of possible clubs is large. In this case, $p_{00}$, the proportion of neither students joining a particular club is less informative than $p_{11}$, the proportion of both students joining that particular club.

\vspace{4mm}\noindent Jaccard similarity and Simpson's overlap coefficient extend the idea of asymmetric measures of co-occurrence and are discussed in further detail in the following sections.

\subsubsection{Jaccard Similarity}

\vspace{4mm}\noindent Jaccard similarity is defined as $J = \frac{p_{11}}{p_{11}+p_{01}+p_{10}}$ and is the proportion that both agent outcomes are $1$ out of all the times either agent outcome is $1$. Jaccard similarity implicitly takes into account asymmetric co-occurrence, assuming that a co-occurring $1$ is more informative than a co-occurring $0$. Because the denominator of the Jaccard similarity is less than 1, Jaccard similarity is always greater than $p_{11}$. Depending the values of $p_{10}$ and $p_{01}$, Jaccard similarity can take on any value greater than $p_{11}$ as long as $p_{11}$ is nonzero.

\vspace{4mm}\noindent Jaccard similarity extends simply using $p_{11}$ as a measure of asymmetric co-occurrence by incorporating information from $p_{10}$ and $p_{11}$. This addresses the cases when either of the variables have a high marginal tendency of being one, and therefore, can result in a relatively high value of $p_{11}$. In the case where one of the variables, say $Y$, in the pair has a high marginal probability of $1$, it may not be desirable to conclude that there is a strong association between the two variables based simply on high ``1" co-occurrence probability since $Y$ is still likely to be $1$ even if $X$ is $0$. Jaccard distance adjusts the measure by reporting the ``1" co-occurrence probability in proportion to when either of the variables is $1$. This will give a relatively lower value of similarity to pairings that have high ``1" co-occurrence probabilities based solely on the fact that one variable is likely to be $1$ all the time.

\vspace{4mm}\noindent In the voting example, imagine that politician $X$ votes ``yes" on a bill regardless of the bill and how everyone else votes. Say for any bill, politician $Y$ has votes ``yes" 70\% of the time. The "yes" co-occurrence proportion is $p_{11}=0.7$. Alternatively, if politicians $A$ and $B$ always vote in the same way, and both vote ``yes" 70\% of the time, then their ``yes" co-occurrence probability is also $p_{11}=0.7$. If only $p_{11}$ is considered as the measure of association, $X$ and $Y$ will be seen as equally similar as $A$ and $B$. Intuitively, however, it may be more reasonable to represent $A$ and $B$ as more similar since their actions do always coincide. The Jaccard similarity between $X$ and $Y$ is $J_{XY} = \frac{0.7}{0.7+0.3}=0.7$, the lowest value it can be with $p_{11}=0.7$, and the Jaccard similarity between $A$ and $B$ is $J_{AB} = \frac{0.7}{0.7} = 1$, the highest value it can be. In this case, Jaccard similarity is able to capture the sense that the association between $A$ and $B$ is stronger than between $X$ and $Y$.

\subsubsection{Simpson's Overlap Coefficient}

\vspace{4mm}\noindent Simpson's overlap coefficient can be defined as $S = \frac{p_{11}}{p_{11}+\min(p_{10}, \text{ }p_{01})}$ and is another extension of the asymmetric measure of association. Simpson's overlap coefficient measures the extent of \textit{overlap} between two variables and is large if either $p_{10}$ or $p_{01}$ is small. Because of the denominator in the Simpson's overlap coefficient expression, it is lower bounded by the Jaccard similarity, so it is also lower bounded by $p_{11}$. Simpson's overlap coefficient for two binary variables is $1$ if one variable is always $1$ whenever the other is a $1$. In the voting example, if politician $X$ will always vote ``yes" on a bill if politician $Y$ votes ``yes", then their Simpson's overlap coefficient will be equal to 1. Politician $X$ can sometimes vote "yes" even if $Y$ votes ``no", so Simpson's overlap coefficient does not necessarily equal Jaccard similarity.

\subsubsection{Summary of Co-occurrence Association Measures}

The parameters given in Table \ref{probs} can be viewed as parameters of the marginal joint probability distribution of $X$ and $Y$, marginalized over all other nodes that may appear in the network. Similarly, marginalized over all other variables, the Jaccard similarity and Simpson's overlap coefficient between $X$ and $Y$ can both be viewed as quantities derived from the marginal conditional distribution. The Jaccard similarity can be stated in probability terms as \\ $J = P(XY=1| X+Y\geq1)$, and Simpson's overlap coefficient can be stated as \\ $S = \max\{P(XY=1|X=1), P(XY=1|Y=1)\}$. 

\vspace{4mm}\noindent A comparison between simple co-occurrence probability, Jaccard similarity and Simpson's overlap coefficient is given in Table \ref{cooccur}. The ordering of the co-occurrence association measures is $0\leq p_{11} \leq J \leq S \leq 1$.

\begin{table}[h!]
\centering
\caption{Comparison of co-occurrence association measures}
\begin{tabular}{|c | c | p{3in}|} 
 \hline
 Description & Expression & Notes\\
 \hline
 Co-occurrence Proportion & $p_{11}$ & Proportion that both variables are $1$\\ 
 \hline
 Jaccard Similarity & $J = \frac{p_{11}}{p_{11}+p_{01}+p_{10}}$ & Accounts for high marginal tendencies; highest when there is no disagreement between the two variables\\
 \hline
 Simpson's Overlap Coefficient & $S = \frac{p_{11}}{p_{11}+\min(p_{10}, \text{ }p_{01})}$ & Accounts for overlaps; highest if one variable is always $1$ when the other is $1$\\
 \hline
\end{tabular}
\label{cooccur}
\end{table}

\subsection{Marginal Independence}

\vspace{4mm}\noindent In addition to co-occurrence, association between two variables can also refer to a notion of independence; whether the realization of one variable reveals any information about the realization of the other. This can be empirically measured through correlation for binary data. Introducing the notions of correlation and independence implicitly assumes a model structure for $X$ and $Y$. Specially, it assumes that $X$ and $Y$ are distributed as bivariate Bernoulli random variables; we can think of the $p_{ij}$'s as joint probability parameters that define the bivariate distribution, and observed proportions as estimates for those parameters. In the politician example, if politicians $X$ and $Y$ are marginally known to vote ``yes" on a bill 40\% of the time, does the fact that $X$ already voted ``yes" change the expectation of a 40\% ``yes" from politician $Y$? If the two vote independently, then the outcome of one would not affect the other. The notion of \textit{marginal independence} can act as a criteria with which to weight edges in a network representation of the data, taking into account every pairwise independence relationship in isolation from all other nodes in the network. 

\vspace{4mm}\noindent For binary variables, one measure of marginal independence is Pearson's correlation coefficient for linear dependence relationships. The correlation coefficient can be written in terms of the probabilities from Table \ref{probs} as $\rho = \frac{p_{11}-(p_{10}+p_{11})(p_{01}+p_{11})}{\sqrt{(p_{10}+p_{11})(1-p_{10}-p_{11})(p_{01}+p_{11})(1-p_{01}-p_{11})}}$.

\vspace{4mm}\noindent The correlation coefficient is signed and satisfies $-1\leq \rho \leq 1$. A correlation coefficient of 0 between two Bernoulli random variables indicates that the random variables are independent, so no additional information can be gained about the outcome of one variable by observing the other. In the binary case, a positive correlation coefficient indicates that the pair of outcomes tend to agree while a negative correlation coefficient indicates that the pair of outcomes tend to disagree. 

\begin{figure}[!htb]
\centering
\caption{Possible correlation coefficients given different values of ``1" co-occurrence probability (top); possible correlation coefficients given different values of Jaccard similarity (bottom)}
  \includegraphics[width=10cm,height=10cm]{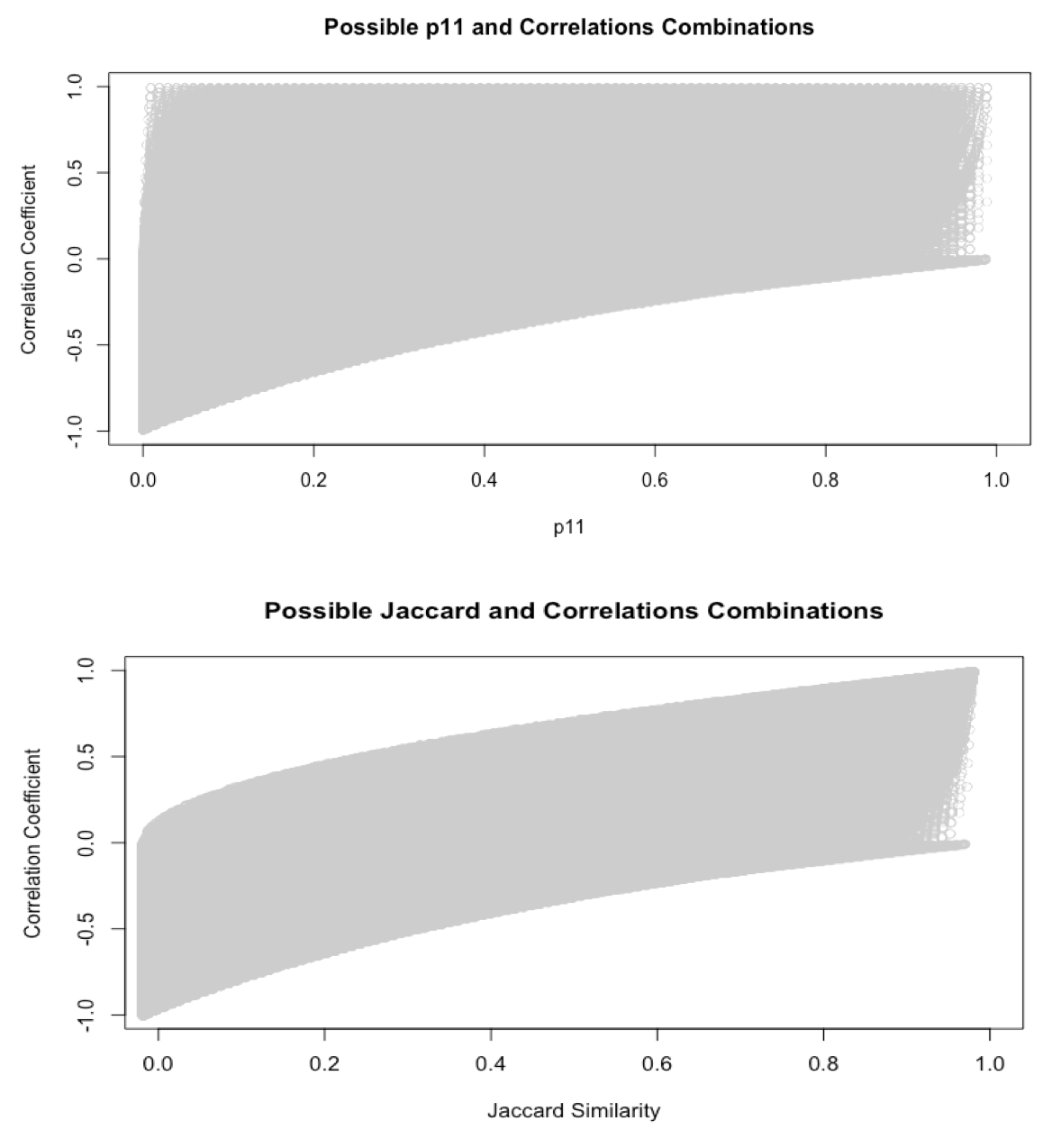}
  \label{jacccor}
\end{figure}

\vspace{4mm}\noindent Correlation differs from co-occurrence measures in that a pair of variables can co-occur very frequently but still have 0 correlation if they are independent. For example, if politicians $X$ and $Y$ both vote ``yes" on a bill 100\% of the time, then they have a co-occurrence probability, Jaccard similarity and Simpson's overlap coefficient of 1 but have a correlation coefficient of 0. This is because knowing how one politician votes in this case does not reveal anything new about how the other politician is likely to vote. Figure \ref{jacccor} (top) shows possible combinations of ``1" co-occurrence probabilities and correlation coefficients. The lower bound for possible correlations is larger for larger values of $p_{11}$, restricting how negative the correlation coefficient can be. The correlation is still able to take on any non-negative value for any given value of $p_{11}$. Figure \ref{jacccor} (bottom) shows possible combinations of Jaccard similarity and correlation coefficients given different configurations of the probabilities from Table \ref{probs}. It can be noted that correlation generally increases with Jaccard similarity, though for a given Jaccard similarity, the correlation coefficient may cover a wide range of positive and negative values. 


\vspace{4mm}\noindent A network based on independence relationships will have the interpretation that outcomes of nodes linked by edges \textit{depend} on each other in a certain way rather than simply proportional to the co-incidence of outcomes.

\section{Model-Based Network Estimation}

\subsection{Dyadic Independence Models}

\vspace{4mm}\noindent If we assume ($X$, $Y$) is distributed according to a bivariate Bernoulli distribution (as in Section 2.2) determined by the probability parameters in Table \ref{probs}, estimating the probability parameters with observed proportions is equivalent to estimating a dyadic independence model, an extension of the Erd\H{o}s-R\'{e}nyi random graph with different edge probabilities \cite{erdos}. Since estimating the probabilities parameters are only concern the outcomes of 2 agents at a time, the model assumes that appearance of each tie is independent of the appearance of every other tie in the network. The edges of dyadic independence models can be viewed as focusing on only the local behaviors of the nodes adjacent to the focal node and disregarding global trends that may be present across the whole network. If every agent in the network is considered at once, each observation of the binary data can be interpreted as a realization of a multivariate Bernoulli distribution\cite{bernoulli}. While pairwise considerations may suffice for building one network and calculating descriptive statistics, if the networks constructed using solely pairwise co-occurence or correlation were viewed as a generative model, the independence assumption between ties is a strong restriction and unrealistic in many real-world social networks. Current research pursuing generative network models are drawn to more flexible and sophisticated models that do not require dyadic independence \cite{amati}, a popular choice of which is the exponential random graph model (ERGM) discussed in the next section.

\subsection{Exponential Random Graph Models}

\vspace{4mm}\noindent While all of the measures of association in all of the previous sections only take into account every pairwise association in isolation from the rest of the nodes in the network, there may be cases where two variables are both affected by the outcome of a third variable but do not directly depend on each other. ERGMs assume partial conditional dependence where the presence of ties between any 2 pairs of nodes in the network may still be conditionally dependent given all other ties in the network \cite{wassermanpat}. In the voting politician example, there may be a leader $A$ whose vote affects how the subordinates $B$ and $C$ vote. This may give $B$ and $C$ a seemingly strong association, whether in terms of co-occurrence or marginal dependence; however, since $B$ and $C$ have no direct effect on each other's outcome, they will not have any measurable association if the outcome of $A$ can be taken into account. The idea of conditional independence attempts to pinpoint, for every variable, the extent of association that can be directly attributed to another variable, after taking into account the outcomes of all other nodes in the network. This can result in a very different network structure from one obtained by estimating co-occurrence or marginal independence relationships. While edges between two nodes in the marginal independence network can include those ``trickle-down" or confounding effects from other nodes in the network, edges drawn from conditional independence measures between two variables represent association that can be directly attributed to each other.

\vspace{4mm}\noindent One way of estimating the strength of association between two variables while conditioning on the outcomes of all other variables is ``leave-one-out" nodewise regression.  Suppose $\boldsymbol{X}$ is a $N \times K$ binary data matrix for $K$ variables of interest and $N$ observations (e.g. $K$ politicians voting ``yes" or ``no" on $N$ bills). Let $X_{-k}$ be the same data matrix with the $k^{th}$ column removed. For each of the $K$ variables, run a regression with that variable as the response and all other variables as explanatory variables. The resulting coefficient vector $\hat{\boldsymbol{\beta}}^k = (\hat{\beta}_0^k,\hat{\beta}_1^k,...,\hat{\beta}_{k-1}^k,\hat{\beta}_{k+1}^k,...,\hat{\beta}_K^k)^T$ from the linear model represent the strength of pairwise conditional association. Since measures of association are generally symmetric, the weight of the edge between the $i^{th}$ and $j^{th}$ variables can be given as $\frac{1}{2}\left(\frac{\hat{\beta}_j^i}{se(\hat{\beta}_j^i)} + \frac{\hat{\beta}_i^j}{se(\hat{\beta}_i^j)}\right)$  where each of the $\hat{\beta}$ estimates are standardized by its standard error to allow for a fairer comparison between variables pairs. Since the data of interest take on binary values, a similar construction can be made using logistic regression instead of linear regression. Additionally, since the primary focus is on the relative association strength between variables conditional on all other variables, this procedure can also be cast as a variable selection problem, and the regression framework can be extended with the LASSO or elastic net to obtain more stable coefficient estimates while also selecting the most significant conditional pairwise associations. Amati et. al. (2018) notes, however, that the unlike the dyadic independence models, the parameter estimates of ERGMs have no intuitive interpretations and even the notion of ``holding other variables constant" can be ambiguous \cite{amati}. Nevertheless, the flexibility of exponential random graph models as an extension to traditional non-model-based network estimation methods makes it increasingly appealing in empirical social science studies \cite{howard}.

\vspace{4mm}\noindent A comparison between Pearson's correlation coefficient and the ``leave-one-out" regression coefficient estimators is given in Table \ref{ind}.

\begin{table}[h!]
\centering
\caption{Comparison of dependence association measures}
\begin{tabular}{|p{1.4in} | c | p{2.5in}|} 
 \hline
 Description & Expression & Notes\\
 \hline
 Pearson's correlation coefficient& $\frac{p_{11}-(p_{10}+p_{11})(p_{01}+p_{11})}{\sqrt{(p_{10}+p_{11})(1-p_{10}-p_{11})(p_{01}+p_{11})(1-p_{01}-p_{11})}}$ & Measures marginal independence; Varies between -1 and 1, negative if variables tend to disagree, positive if variables tend to agree, 0 if variables are marginally independent\\ 
 \hline
 Average ``leave-one-out" regression coefficient & $\frac{1}{2}\left(\frac{\hat{\beta}_j^i}{se(\hat{\beta}_j^i)} + \frac{\hat{\beta}_i^j}{se(\hat{\beta}_i^j)}\right)$ & Measures conditional independence between a pair of variables while accounting for all variables; sign indicates direction of association and absolute magnitude indicate strength of association\\
 \hline
\end{tabular}
\label{ind}
\end{table}

\section{New York Tech Membership Data Set}

\vspace{4mm}\noindent To illustrate the differences between networks constructed using the various measures, the following section compares networks constructed using each measure of association from a data set of online group memberships from a community of people. The membership data is of people on the social networking platform ``Meetup.com" where individuals join ``meetup" groups to facilitate special interest activities among people in a certain area; the data set used involves 2009 membership in meetup groups in the New York metropolitan area that have been joined by at least 2 people who are also in the New York Tech Meetup group. The nodes that form the network of interest are the meetup groups that are represented in the data set, and associations between the groups are based on memberships of the New York Tech Meetup members. Because meetup groups that are joined by members of New York Tech Meetup often focus on special interests geared towards New York's increasing focus on technology, the networks themselves are of interest from a economic and sociologic perspective as it captures the organizational structure of technological specialties in the community.

\vspace{4mm}\noindent The groups in the data set ranges from meetups that specialize in coding in various programming languages, to business/startup networking gatherings, to social and recreational activities. The relationship between the list of all meetup groups and all individuals of interest can be encoded in a binary matrix $X$, where each row corresponds to an individual and each column corresponds to a meetup group. The element $X_{ij}$ is a 1 if the $i^{th}$ individual is a member of the $j^{th}$ meetup group, and 0 otherwise. A simple representation of the data in matrix form is given below.

\vspace{4mm}\noindent Suppose member A is part of the groups \{Software Development , Big Data, Learn C++, Poker Nights\}; member B is part of \{Technology Meetup, Learn C++, Software Development\}; member C is in \{Web Designing, Poker Nights, Local Musicians\}; member D is in \{Poker Nights, Local Music, Jazz Band\}. The data matrix representing this set of members and specialized groups is 

\[
\begin{array}{c}
   \text{\scriptsize\sl \qquad\qquad\quad
    Software Dev ~ Big Data ~ C++ ~ Tech Meetup ~ Web Design ~ Poker ~ Musician ~ Jazz} 
\\
X=
\begin{array}{cc}
{\begin{array}{c} A \\ B \\ C \\ D \end{array} }
& 
{\left[ 
\begin{array}{cccccccc}
\quad\qquad
1\quad &\quad1\quad&\quad1\quad&\quad0\quad&\quad0\quad&\quad1\quad&\quad0\quad&\quad0 \quad\\
\qquad 1&0&1&1&0&0&0&0 \\
\qquad 0&0&0&0&1&1&1&0 \\
\qquad 0&0&0&0&0&1&1&1 \\
\end{array}
\right]}
\end{array}
\end{array}
\]
  
\begin{figure}[!htb]
\centering
\caption{Comparison of weights given to each pair of groups by different association measures for Meetup groups data}
  \includegraphics[width=17.5cm,height=9.5cm]{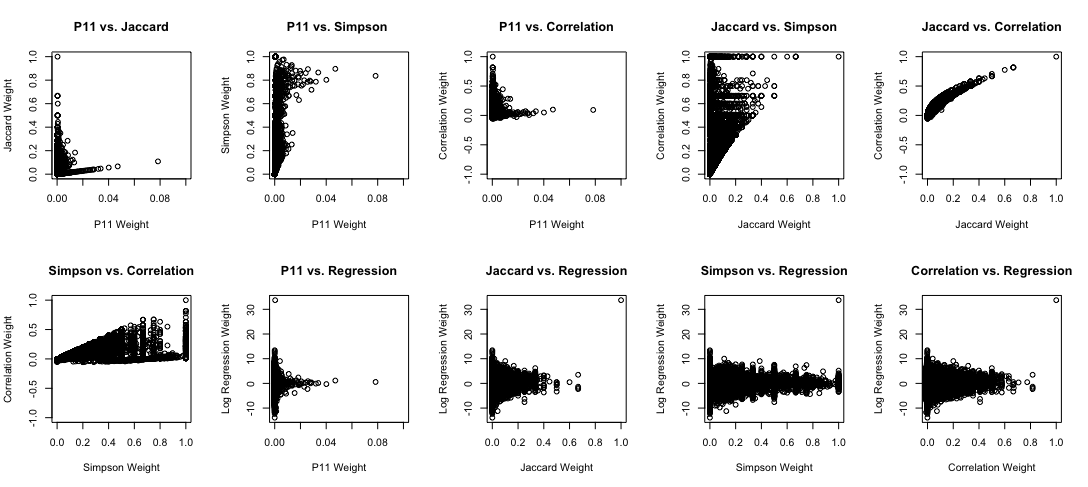}
  \label{pairmeasures}
\end{figure}

\vspace{4mm}\noindent The 2009 New York Tech Meetup membership data contains 859 Meetup groups across 5,883 members. Each person's membership in the dataset is regarded as an independent and identically distributed realization from an underlying 589-variate Bernoulli distribution. The Bernoulli variables in this example represent each group, and is a $1$ for a given member if the member is part of the group, and is a $0$ otherwise. Using this data, the co-occurrence probability, $p_{11}$, of two Meetup groups is estimated by the number of members who are part of both groups divided by 5,883. The other probabilities can be estimated in a similar way, and the co-occurrence and correlation measures of association can be calculated accordingly. Figure \ref{pairmeasures} shows the weight of every pair of Meetup groups as measured by the different association measures. For the conditional independence regression measure, the average standardized coefficients from corresponding linear regressions are taken (the quantity given by Table \ref{ind}). Since the standardized regression coefficients do not fall in a fixed range like the other association measures, the logarithmic magnitude of the quantities used for ease of visualization (the signs are left as is).

\vspace{4mm}\noindent Every point in each plot of Figure \ref{pairmeasures} represent a pair of Meetup groups, and the relative position of the point in each point is determined by the weight that the particular measure of association gives to that pair of groups based on membership data. The points in the ``P11 vs. Jaccard" and ``Jaccard vs. Correlation" plots all fall in the respective feasible regions given in Figure \ref{jacccor}. The relationship $p_{11}\leq J\leq S$ is captured by the ``P11 vs. Jaccard", ``P11 vs. Simpson" and ``Jaccard vs. Simpson" plots as all of the points lie above the 45 degree identity line. Correlation appears to generally increase with $p_{11}$, Jaccard similarity as well as Simpson's overlap coefficient, and the relationship is strongest with Jaccard similarity. The log standardized regression coefficients, on the other hand, show no clear pattern with any of the measures of association, suggesting that while co-occurrence and marginal independence captures some of the same signals in the data, the regression method for measure conditional independence does not pick up on the same set of signals as the other measures. 

\vspace{4mm}\noindent Table \ref{cooccuredges} lists the 5 edges with the largest weight according to the three co-occurrence association measures. Table \ref{indedges} lists the 5 edges with the largest magnitude of weight according to the independence measures. 

    \begin{table}
        \centering
        \caption{Strongest Edges from Co-occurrence measures}
        \label{cooccuredges}
        \begin{tabular}{|c||p{2.5in}|p{2.5in}||c||}
        \hline
         & \multicolumn{2}{|c||}{$p_{11}$ Co-occurrence Probability} & Weight \\
         \hline
          1. &NYC Entrepreneur Meetup & NYC Business Networking Group & 0.014\\
          2. &NY Video & NY Venture Collaboration & 0.013\\
          3. &BigScreen LittleScreen NYC & NY Video & 0.012\\
          4. &NY Venture Collaboration & NYC Entrepreneur Meetup & 0.011\\
          5. &NY Video & Brooklyn Futurist Meetup & 0.010\\
            \hline
            \hline
           & \multicolumn{2}{|c||}{Jaccard Similarity} & Weight\\
           \hline
           1. & Chinese \& Taiwanese Chat & Flushing Business Meetup & 1.00\\
           2. & Bodhi Lounge & Get Your Dance On NYC & 0.67\\
           3. & NYC Baby Boomers Meetup & NYC Singles Meetup Club & 0.67\\
           4. & Long Island Meetup and Go & The Science of Well-Being & 0.67\\
           5. & NYC Social Singles & Westchester's Singles Group & 0.067\\ 
           \hline
           \hline
           & \multicolumn{2}{|c||}{Simpson's Overlap Coefficient} & Weight\\
           \hline
           1. & NY Social Society & Chinese \& Taiwanese Chat & 1.00\\
           2. & NY Social Society & Flushing Business Meetup & 1.00\\
           3. & NY Social Society & New York 80's Music Meetup & 1.00\\
           4. & NY Social Society & OneTaste New York & 1.00\\
           5. & NY Social Society & Hospitality Industry Social Group & 1.00\\ 
	  \hline
        \end{tabular}
    \end{table}
    
    \begin{table}
        \centering
        \caption{Strongest Edges from Independence measures}
        \label{indedges}
        \begin{tabular}{|c||p{2.5in}|p{2.5in}||c||}
        \hline
         & \multicolumn{2}{|c||}{Pearson's Correlation Coefficient} & Weight \\
         \hline
          1. & Chinese \& Taiwanese Chat & Flushing Business Meetup & 1.00\\
           2. & Bodhi Lounge & Get Your Dance On NYC & 0.82\\
           3. & NYC Baby Boomers Meetup & NYC Singles Meetup Club & 0.82\\
           4. & Long Island Meetup and Go & The Science of Well-Being & 0.82\\
           5. & NYC Social Singles & Westchester's Singles Group & 0.82\\ 
            \hline
            \hline
           & \multicolumn{2}{|c||}{Regression Coefficient for Conditional Independence} & Weight\\
           \hline
           1. & Chinese \& Taiwanese Chat & Flushing Business Meetup & $4.27\times 10^{14}$\\
           2. & ``Amazing Singles" & Greenwich-Westchester Singles & $4.84\times10^{2}$\\
           3. & Minority Wall Street & Westchester Tennis Meetup & $1.81\times10^{2}$\\
           4. & Mix, Mingle, Connect & Westchester Film and TV & $1.80\times 10^2$\\
           5. & NYC World of Warcraft & SEO Super Power & $1.70\times 10^2$\\ 
           \hline
           
        \end{tabular}
    \end{table}
    
\vspace{4mm}\noindent In accordance with Figure \ref{pairmeasures}, Tables \ref{cooccuredges} and \ref{indedges} show that different measures of association emphasize different aspects of the ties between nodes, which can create a wide range of network representations of the same data set. For the co-occurrence probabilities, $p_{11}$ gives more weight to the overall more popular groups while Jaccard and Simpson's measures of association both normalize by the size of the group. This is shown by the strongest weights for $p_{11}$ being all among professional and technological focused while the Jaccard and Simpson's weights pick out social and leisurely activities. Since the target population of the data set is centered on people who have also joined New York Tech Meetup, professional interest groups such as those focused on entrepreneurship, business networking and venture collaboration should also be highly popular among the members, given them high estimated co-occurrence probabilities between one another. The social groups range widely in terms of focus and participation such that no single social meetup group is as comparably popular as the professional groups. Jaccard similarity and Simpson's overlap coefficient, however, normalize the weights by a factor determined by $p_{11}, p_{01}$ and $p_{10}$, which allows the social groups to be up-weighted. It can also be noted that Simpson's overlap coefficient gives the largest weight to pairs where one group is wholly contained in another group so the top 5 edges all have weights of 1. The NY Social Society is likely a large general organization with members that compose smaller breakout groups that focus on more specific social activities. Members that part of the smaller, more focused groups are likely also part of the larger, more general group, giving those ties relative large weights by Simpson's overlap measure. The 5 edges with the largest weight in terms of correlation is the same 5 edges with the largest Jaccard similarity measure. This is not surprising given that the two measures for this data set are shown to have a strong relationship with each other in Figure \ref{pairmeasures}. Overall, depending on the type of relationship between nodes desired for further analysis, the choice of association measure can greatly affect the structure of the estimated neteork. 

\section{Discussion}

\vspace{4mm}\noindent The results in Section 4 applying the various association measures to estimate a network structure using a real set of data shows that the choice of association measure can give very different interpretations to the ties between nodes of the network. The occurrence measures are more intuitive and have simple expressions in terms of estimated probabilities. For the New York Tech Meetup data set, edges weighted on $p_{11}$ are larger if the incident nodes with high overall popularity. Jaccard similarity weights a pair of binary variables highly if it is very unlikely that the outcomes disagree with each other. For the New York Tech Meetup data, this manifests in groups where members are likely to either join both or neither. This does not require the groups to be very popular overall, which is reflected in the top edges composed of groups that are relatively modest in size. However, if the analysis desired from the network representation should emphasize overall popular groups, pre-processing of the data where only groups above a high threshold of membership can focus the sample on the population of greater interest. Simpson's overlap coefficient gives the highest weights to associations where one variable encompass the other. Unlike Jaccard similarity, in the NY Tech example, Simpson's overlap weight is high if being a member of one group signifies that it is very likely to also be a member of another group, but the reverse need not necessarily be true. 

\vspace{4mm}\noindent Correlation in the binary case can be a measure of marginal independence and measures whether knowing a outcome of one variable gives any extra information on the outcome of another. If knowing the outcome of one variable is highly informative of the outcome of another, the strength of the association between the two variables will be high by the correlation measure. A higher correlation measure is typically associated with a higher Jaccard similarity as shown in Figure \ref{jacccor}, and this is the case in the NY Tech Meetup example. Unlike the co-occurrence measures, the correlation measure is directional; the sign of the association reveals the whether the outcomes of the variable pair tend to agree or disagree in the binary case. Correlation is a measure of marginal independence, considering each pair of variables in isolation from all other variables on the network. Conditional independence is also important to consider when assessing the strength of association between variables. Knowing the outcome of one variable may only reveal information about the outcome of another variable insofar as their mutual relationship to a third variable. Conditional independence measures try to filter out the association of a pair of variables that is directly attributed to each other, taking into account all other variables on the network. The ``leave-one-out" regression procedure discussed in this paper is one method to obtain a measure of conditional independence, and since the regression framework is easily extended to various other situations, the procedure is a highly versatile measure for conditional independence. From the NY Tech Meetup example, conditional independence appears to capture a completely different set of signals from the data compared to the co-occurrence and correlation measures of marginal association.

 \vspace{4mm}\noindent As network analysis become an increasingly popular way to visualize and analyze data, especially binary data, the choice of how to estimate the network structure from data becomes increasingly important. Because of the versatility of network visualization, a vague concept of what the network represents can hinder useful interpretations of the results. This paper presents several possible ways of defining a network, each representing a subtly different interpretation of the underlying data. Different ideas of association are shown to result in very different network results, and any empirical analysis should be conducted only after careful consideration of the desired interpretation of the network.


\begin{thebibliography}{99}

\bibitem{amati} Amati, V., Lomi, A., Mira A. (2018). ``Social Network Modeling". \textit{Annual Review of Statistics and Applications}. 5:343-369. 

\bibitem{aral} Aral, S., Muchnick, L., Sundararajan, A. (2009). ``Distinguishing Influence-Based Contagion from Homophily-Driven Diffusion in Dynamic Networks." \textit{Proceedings of the National Academy of Sciences}. 106(51): 21544-21549.

\bibitem{bernoulli} Dai, B., Ding, S., Wahba, G. (2013). ``Multivariate Bernoulli Distribution". \textit{Bernoulli}. 19(4):1465-1483.

\bibitem{borgatti} Borgatti, S.P., Mehra, A., Brass, D.J., Labianca, G. (2009). ``Network Analysis in the Social Sciences." \textit{Science}. 323(5916): 892-895.

\bibitem{dellaposta} DellaPosta D. (2017). ``Network Clousure and Integration in the Mid-20th Century American Mafia." \textit{Social Networks}. 51: 148-157.

\bibitem{erdos} Erd\H{o}s, P., R\'{e}nyi, A. (1960). ``On the Evolution of Random Graphs". \textit{Publications of the Mathematical Institute of the Hungarian Academy of Sciences}, 5: 17-61.  

\bibitem{kartik} Hosanagar, K., Fleder, D., Lee, D., Buja, A. (2014). ``Will the Global Village Fracture into Tribes? Recommender Systems and their Effects on Consumer Fragmentation." \textit{Management Science}. 60(4): 805-823.

\bibitem{kadushin} Kadushin, C. (2012). \textit{Understanding Social Networks: Theories, Concepts, and Findings}. New York: Oxford University Press. 

\bibitem{howard} Kim, J., Howard, M., Cox Pahnke, E., Boeker, W. (2016). ``Understanding Network Formation in Strategy Research: Exponential Random Graph Models." \textit{Strategic Management Journal}. 37: 22-44. 

\bibitem{newman} Newman, M.E.J. (2010). \textit{Networks: An Introduction}. Oxford, United Kingdom: Oxford University Press. 

\bibitem{porter} Porter, M.A., Mucha, P.J., Newman, M.E.J., Warmbrand, C.M. (2005). ``A network analysis of committees in the U.S. House of Representatives." \textit{Proceedings of the National Academy of Sciences}. 102(20): 7057-7062.

\bibitem{susarla} Susarla, A., Oh, J., Tan, Y. (2012). ``Social Networks and the Diffusion of User-Generated Content: Evidence from YouTube." \textit{Information Systems Research}. 23(1): 23-41.

\bibitem{wassermanpat} Wasserman S., Pattison P. (1996). ``Logit Models and Logistic Regressions for Social Networks: I. An Intorduction to Markov Graphs and $p^*$". \textit{Psychometrika}, 61(3): 401-425.

\end{thebibliography}
\end{document}